\documentclass[final, 3p, times, twocolumn]{elsarticle}

\usepackage{lineno,hyperref}
\modulolinenumbers[5]

\journal{Journal of \LaTeX\ Templates}

\begin{document}

\begin{frontmatter}

\title{Development of a Neutron Imaging Sensor using INTPIX4-SOI Pixelated Silicon Devices}


\author[icepp]{Y. Kamiya\corref{corresp}}
\ead{kamiya@icepp.s.u-tokyo.ac.jp}
\cortext[corresp]{Corresponding author}

\author[KEK1]{T. Miyoshi}
\author[KEK2]{H. Iwase}
\author[icepp]{T. Inada}
\author[Tokyo1]{A. Mizushima}
\author[Tokyo1]{Y. Mita}
\author[Tokyo2]{K. Shimazoe}
\author[KUR]{H. Tanaka}
\author[KEK3]{I. Kurachi}
\author[KEK1]{Y. Arai}

\address[icepp]{Department of Physics and International Center for Elementary Particle Physics,
The University of Tokyo, Tokyo 113-0033, Japan}
\address[KEK1]{Institute of Particle and Nuclear Studies, 
High Energy Accelerator Research Organization (KEK), Ibaraki 305-0801, Japan}
\address[KEK2]{Radiation Science Center, 
High Energy Accelerator Research Organization (KEK), Ibaraki 305-0801, Japan}
\address[Tokyo1]{Department of Electrical Engineering and Information Systems, 
The University of Tokyo, Tokyo 113-8656, Japan}
\address[Tokyo2]{Department of Nuclear Engineering and Management, 
The University of Tokyo, Tokyo 113-0033, Japan}
\address[KUR]{Institute for Integrated Radiation and Nuclear Science, 
Kyoto University, Osaka 590-0458, Japan}
\address[KEK3]{Department of Advanced Accelerator Technologies, 
High Energy Accelerator Research Organization (KEK), Ibaraki 305-0801, Japan}

\begin{abstract}
We have developed a neutron imaging sensor 
based on an INTPIX4-SOI pixelated silicon device. 
Neutron irradiation tests are performed at several neutron facilities to investigate sensor's responses for neutrons.
Detection efficiency is measured to be around $1.5$\% for thermal neutrons.
Upper bound of spatial resolution is evaluated to be $4.1 \pm 0.2 ~\mu$m 
in terms of a standard deviation of the line spread function.
\end{abstract}


\end{frontmatter}

\section{Introduction}

Imaging sensors based on semiconductor technologies are widely utilized 
for optical and quantum beam imaging,
because of their superior spatial and temporal resolutions, 
and their high degree of freedom in handling due to availability of electrical control.
Recently, several developments of neutron imaging sensors with pixelated devices
have been reported and are
aimed at imaging ultra-cold neutrons (UCNs).
The reported sensors are the Timepix-based ($55 ~\mu$m pixel size) with $^{6}$LiF coating \cite{timepix1, timepix2},
Hamamatsu back-illuminating CCD-based ($24 ~\mu$m pixel size) with $^{10}$B coating \cite{hamaTokyo}, 
Logitech Webcam CMOS-based ($3 ~\mu$m pixel size) with $^{6}$LiF and $^{10}$B layers \cite{CMOS-Webcam}, 
Hamamatsu back-illuminating CCD-based ($14 ~\mu$m pixel size) with $^{10}$B coating \cite{hamaGranit}, 
and the DECam CCD-based ($15 ~\mu$m pixel size) with $^{10}$B coating \cite{DECam}.
Those technologies are expected to be useful also for industrial applications 
such as a non-destructive imaging of light elements \cite{NeuRadiography1, NeuRadiography2}.

Spatial resolution for the Timepix-based and the CCD-based sensors were
evaluated to be $2.3 ~\mu$m in terms of a standard deviation of the point spread function (PSF) \cite{timepix1} 
and $3.4 ~\mu$m in the line spread function (LSF) \cite{hamaTokyo}, respectively. 
For the Webcam CMOS-based sensor, upper bound of the resolution was estimated to be $60 ~\mu$m.
These sensors have a detection efficiency of several tens percents and more for UCNs,
however, it is only a few percents or less for cold and thermal neutrons.
Therefore, even at the expense of its spatial resolution,
neutron sensors with a micro-structured converter 
aiming for higher detection efficiency have been developed \cite{TrenchNeutron1, TrenchNeutron2}.

In this paper, we report developments of new neutron imaging sensors that are based on 
the INTPIX4-SOI pixelated silicon device \cite{INTPIX4_1,INTPIX4_2},
aiming at a fine spatial resolution with time-resolving readout.
In section 2 the INTPIX4 chip is introduced, 
followed by the principal of operation and sensor design in section 3.
Based on measurements from thermal neutron facilities, 
event identification processes using the shapes of a charge cluster are discussed in section 4.
Detection efficiency was estimated under the criteria that we decided to use.
We also evaluated the spatial resolution
by fitting edges of shadow of a neutron mask, which has a fine structure with sub-micron accuracy. 
The detail of the measurement is presented in section 5.
In contrast with our previously developed sensor \cite{hamaTokyo}, 
its faster readout time helps us to measure a neutron event with resolution in $O(10^{-2})$ seconds or less. 
The SOI-CMOS architecture with on-chip-circuit-layers is attractive, 
because of its availability of integrations of some rich functions on chip.
We are planning to integrate, for example, a dedicated self-triggering circuit with a cluster identification engine.
The discussion here will be the first step in the development of these new sensors.

\section{INTPIX4}

\begin{figure}[htb]
\begin{center}
\includegraphics[width=0.85\linewidth]{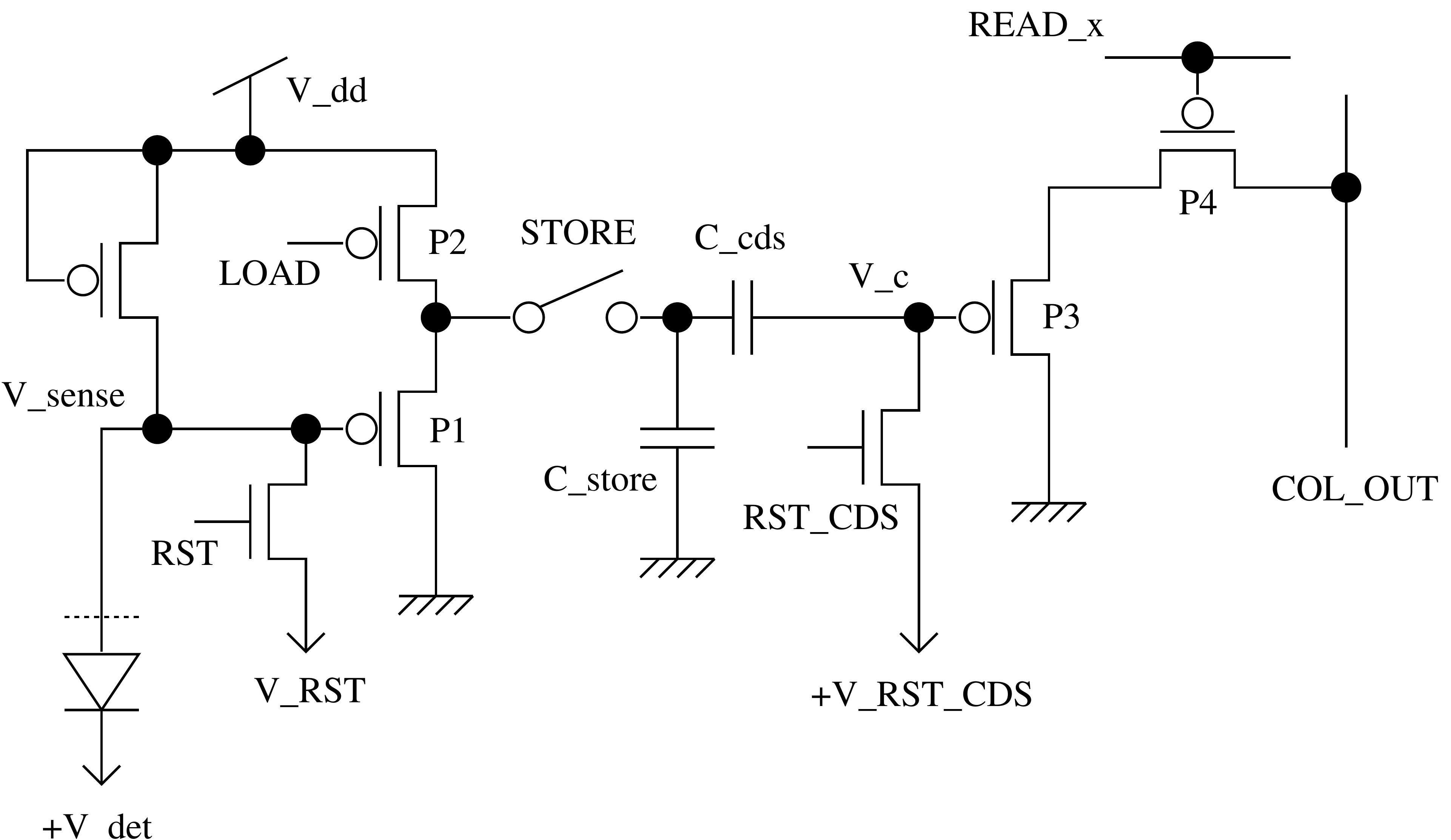}
\caption{Front-end circuit embedded on a SOI wafer.}
\label{circuit}
\end{center}
\end{figure}

INTPIX4 is an charge integrating device 
of $17 ~\mu$m $\times$ $17 ~\mu$m pixels with circuit matrix on a ``silicon on insulator" (SOI) wafer \cite{SOI}, 
in which correlated double sampling circuits and storage capacitors are implemented 
for synchronized operation with external triggers. 
Figure \ref{circuit} shows the front-end circuit,
which is embedded on the circuit layer of SOI.
The number of pixels is $832$ $\times$ $512$ and 
active area is $14.1$ mm $\times$ $8.7$ mm.
It is divided to 13 blocks of 
$64$ $\times$ $512$ pixels and the charges can be read out parallelly for each block.
Readout time is around $280$ ns/pixel, which allows
a maximum frame rate to be around 100 Hz.
Float-Zone (FZ) silicon wafer of $300 ~\mu$m thickness is used for the sensor's substrate.
Its resistivity is a few kOhm. 
The circuit gain was evaluated to be $13.1 ~\mu$V/$e^{-}$ \cite{Kanazawa}.
Al layer of $200$ nm is 
coated on back side of the sensor in the fabrication process.

The INTPIX4 chip is wire-bonded to a sensor board and
the sensor board is connected to a SEABAS2-board, a general-purpose readout platform, developed in High Energy Accelerator Research Organization (KEK) \cite{INTPIX4_2}.
The SEABAS2-board has a Virtex5 FPGA which controls the sensor circuit. 
It also has a Virtex4 FPGA \cite{SiTCP} that allows data transfers over the TCP/IP protocol.

\section{Fabrication of Neutron-sensitive Sensor}

Recently, semiconductors containing $^6$Li
such as $^6$LiInSe$_2$ or $^6$LiInP$_2$Se$_6$ have been reported \cite{LiInSe_1,LiInSe_2,LiInSe_3}.
Pixelated read-out feature is not implemented in these materials yet, therefore,
the conventional scheme, making a neutron-to-charged-particle conversion layers 
formed in front of the pixelated devices, 
is currently the best choice for achieving neutron-sensitive imaging sensors.

The following nuclear reactions for the neutron conversion processes are utilized,
\begin{eqnarray}
n + ^{10}\!{\rm B} &\rightarrow& \alpha_{(1.47 ~{\rm MeV})} + ^7{\rm Li}_{(0.84 ~{\rm MeV})} + \gamma_{(0.48 ~{\rm MeV})},\\
n + ^{10}\!{\rm B} &\rightarrow& \alpha_{(1.78 ~{\rm MeV})} + ^7{\rm Li}_{(1.01 ~{\rm MeV})},
\end{eqnarray}
where the secondary particles are emitted in back-to-back directions.
Branching ratios are $94$ \% and $6$ \%, respectively. 
One of the emitted particles enters the active volume of the sensor,
and stops with corresponding energy loss rate.

$O(10^5)$ electron-hole pairs are generated along the traveling pass of the secondary particle
and the holes start to drift to the electrodes.
In the very early phase during the drifting, 
charges are repulsed each other by the Coulomb force 
due to its relatively high charge density, 
and then the natural diffusion process follows.
Because of these diffusion processes, 
the charges are shared by multiple pixels 
and the cluster's shape becomes a symmetrical two-dimensional Gaussian or slightly bell-shaped distribution.

\begin{figure}[htb]
\begin{center}
\includegraphics[width=0.85\linewidth]{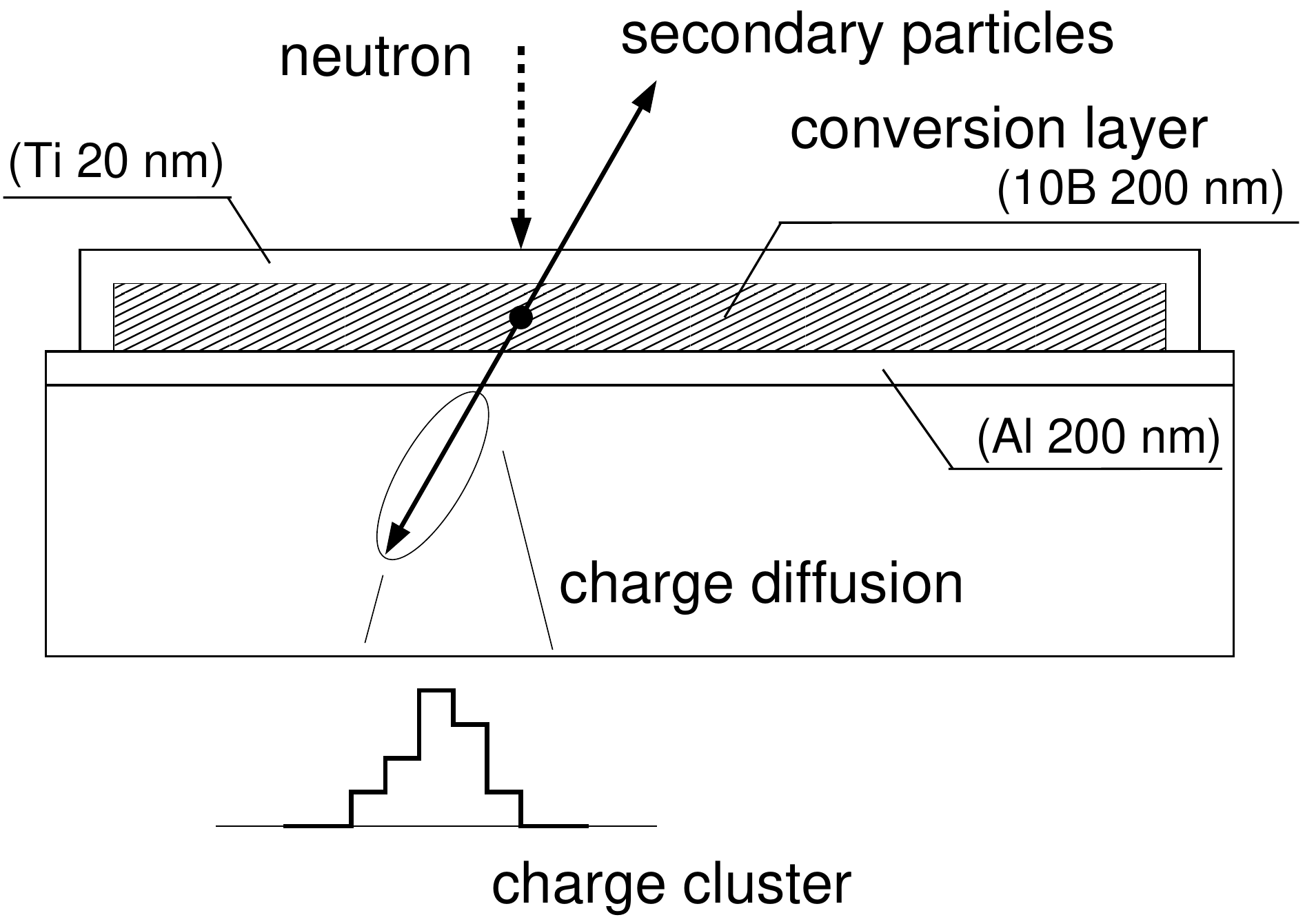}
\caption{Cross-section view of the sensor.
$^{10}$B and Ti layers were deposited by physical vapor deposition processes.
$^{10}$B nucleus absorbs a neutron and emits charged secondary particles in back-to-back directions.
One of them deposits its energy in the active volume of the INTPIX4 device.
Generated holes drift to pixelated electrodes
and diffuse two-dimensionally to the x-y plane in the process.}
\label{sensor}
\end{center}
\end{figure}

Figure \ref{sensor} shows schematic drawing of the sensor and the conversion layer 
in the cross-section view.
The $^{10}$B conversion layer was formed directly on the backside Al layer 
using the argon RF sputtering technique.
The thickness of this layer was adjusted to be 200 nm.
The $^{10}$B layer was covered with thin Ti film 
to prevent oxidation, and degradation by chipping.

We used the Shibaura CFS-4EP-LL sputtering machine at the University of Tokyo, 
maintained under the framework of the Nanotechnology Platform program in Japan.
Sputtering sources are 3-inch disks, located at 110 mm distance from the samples.
Target and substrate are held vertically 
and side-by-side aligned to avoid contamination from residual particles which are falling on.
There are a pre-vacuum chamber and a load lock system,
which can help us to keep the main chamber vacuum even when changing sample configurations.
During the sputtering, the sample holder disk rotates with 20 rpm speed.
Sputtering rate for $^{10}$B is 0.03 nm/s with 400 W RF power in 470 mPa Ar,
then the $^{10}$B layer of 200 nm thickness is formed in around 2 hours.

\section{Neutron Events}

\begin{figure}[htb]
\begin{center}
\includegraphics[width=0.85\linewidth]{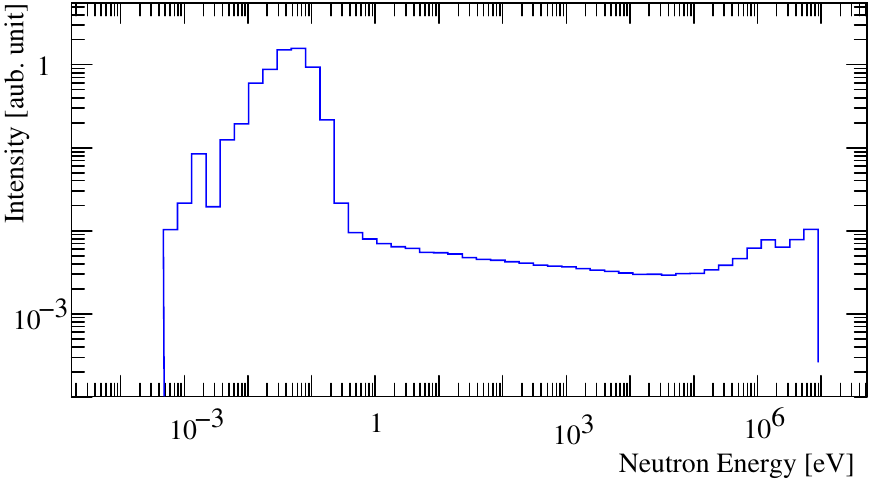}
\caption{Expected energy distribution simulated by PHITS \cite{PHITS}.
Most of neutrons are in the range of $10^{-2}$ eV to $10^{-1}$ eV.}
\label{STN}
\end{center}
\end{figure}

A $^{10}$B coated INTPIX4 ($^{10}$B-INTPIX4) sensor was tested with thermal neutrons 
supplied from the Standard Thermal Neutron Irradiation Laboratory at KEK.
Neutrons are emitted from a $^{241}$Am-Be neutron source with $37 $ GBq intensity,
which is located in the center 
of a $1.9$ m (width) $\times$ $2.5$ m (length) $\times$ $1.9$ m (hight)
carbon pile (density is 1.81 g/cm$^3$).
Neutron intensity is measured to be 20 /cm$^2$/s 
at the point of our testing sensors that is 1.25 m horizontally away from the neutron source.
The expected energy distribution is simulated by PHITS \cite{PHITS}, as shown in Fig. \ref{STN}.
Totally, $2.6877 \times 10^{6}$ images were taken.

Events are selected with the following scheme. 
\begin{enumerate}
    \item estimate pedestals of each pixel on each image by averaging successive 10 images before and after the image to be evaluated, and make the pedestal corrections.
    \item find the pixel which has higher charge than the threshold level which corresponds to a 34 keV energy deposit.
    \item check the adjacent pixels inside $7 \times 7$ distance to find a local maximum pixel, and define it as a seed of cluster.
    \item determine a $7 \times 7$ pixels frame centered on the seed. Any structures inside the frame is treated as one event/cluster.
    \item if you find the only one pixel, which have charge higher than 3.4 keV (10\% of the threshold), in the frame, 
    this event is due to the noisy pixel, then reject it.
\end{enumerate}

The sum of the pixels' charges in the frame ($Q$)
corresponds to the total energy deposit of the secondary particle.
The charge weighted mean of the positions of each pixel ($\vec{\mu} = (\mu_x, \mu_y)$)
represents an estimate of neutron incident position.
To identify neutron events from electron-like non-symmetric events or noisy images affected by a common noise,
we use the 2nd order center moment on x-axis and y-axis ($\vec{\sigma^2} = (\sigma_x^2, \sigma_y^2)$)
as additional identifiers.
They are written as
\begin{eqnarray}
    Q &=& \sum_{i=1}^{i=49} q_i, \\
    \vec{\mu} &=& \frac{1}{Q} \sum_{i=1}^{i=49} q_i \vec{r}_i\\
    \vec{\sigma_k^2} &=& \frac{1}{Q} \sum_{i=1}^{i=49} q_i (\vec{r}_i - \vec{\mu})^2, 
\end{eqnarray}
here $q_i$ is a charge of $i$-th pixel in the frame, 
and $\vec{r}_i = (x_i, y_i)$ is a pixel position.

\begin{figure}[htb]
\begin{center}
\includegraphics[width=0.95\linewidth]{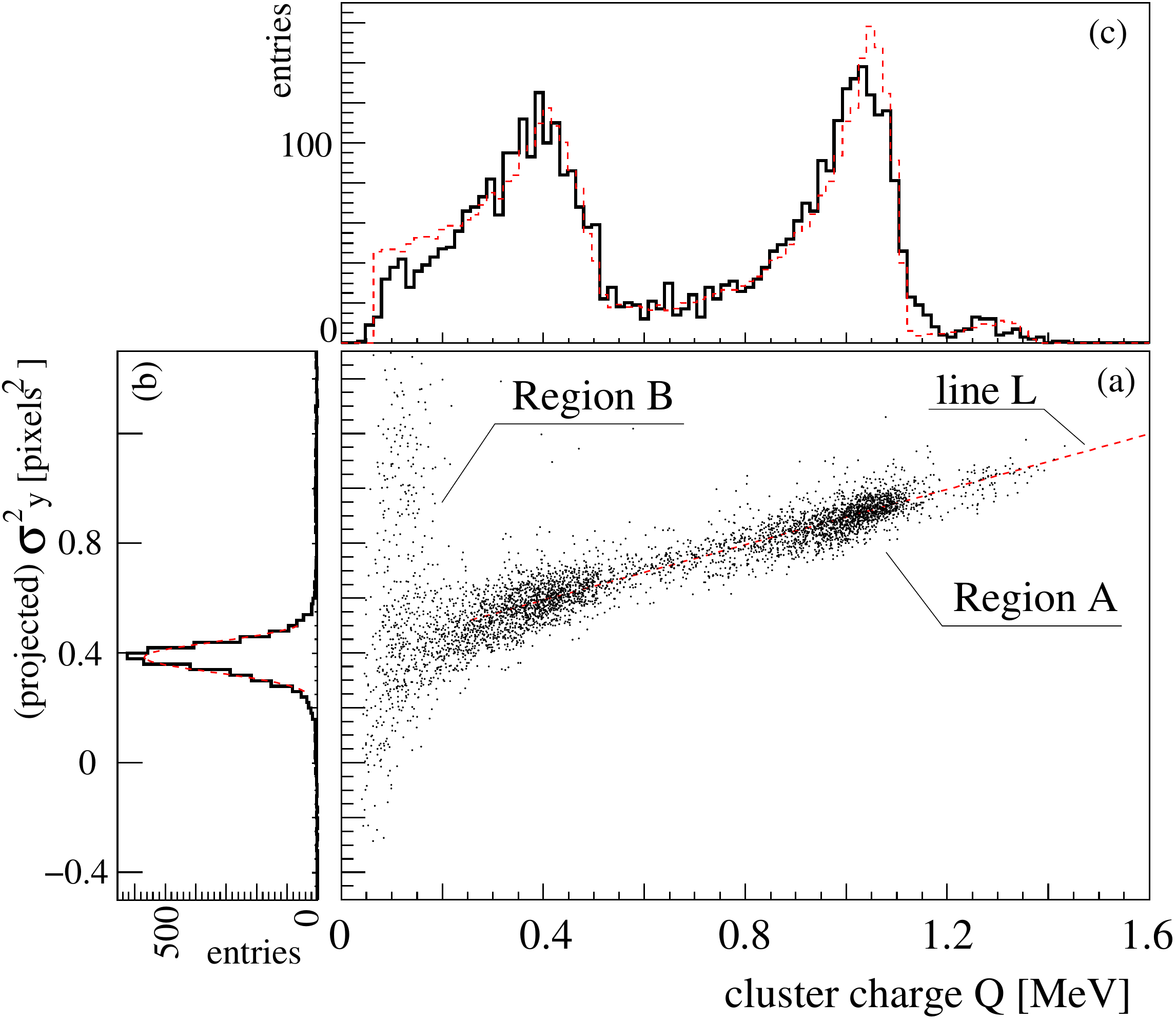}
\caption{
(a) Scatter plot for events in $\sigma_y^2$ vs $Q$ plane.
A dashed line (the Line L) is evaluated to be
$\sigma_y^2 = \hat{a}Q + \hat{b}$, where $ \hat{a} = 5.0 \times 10^{-7} $ (pixel)${}^2/$eV and $\hat{b} =  0.39$ (pixel)${}^2$,
by fitting the distribution in $Q > 0.25$ MeV.
(b) Projected $\sigma_y^2$ distribution in $Q > 0.25$ MeV range 
along the line L. Mean and a standard deviation of the best fit Gaussian are 0.39 and 0.053, respectively.
(c) Cluster charge $Q$ distribution in the criteria of neutron events,
$|\sigma_y^2 - \hat{a}Q - \hat{b}| < 0.16$.
Dashed histogram shows the best fit distribution with an inactive layer just under the Al skin of the INTPIX4 device.
A thickness of the inactive layer is estimated to be $220 \pm 20$ nm.
}
\label{scatt1}
\end{center}
\end{figure}
\begin{figure}[htb]
\begin{center}
\includegraphics[width=0.75\linewidth]{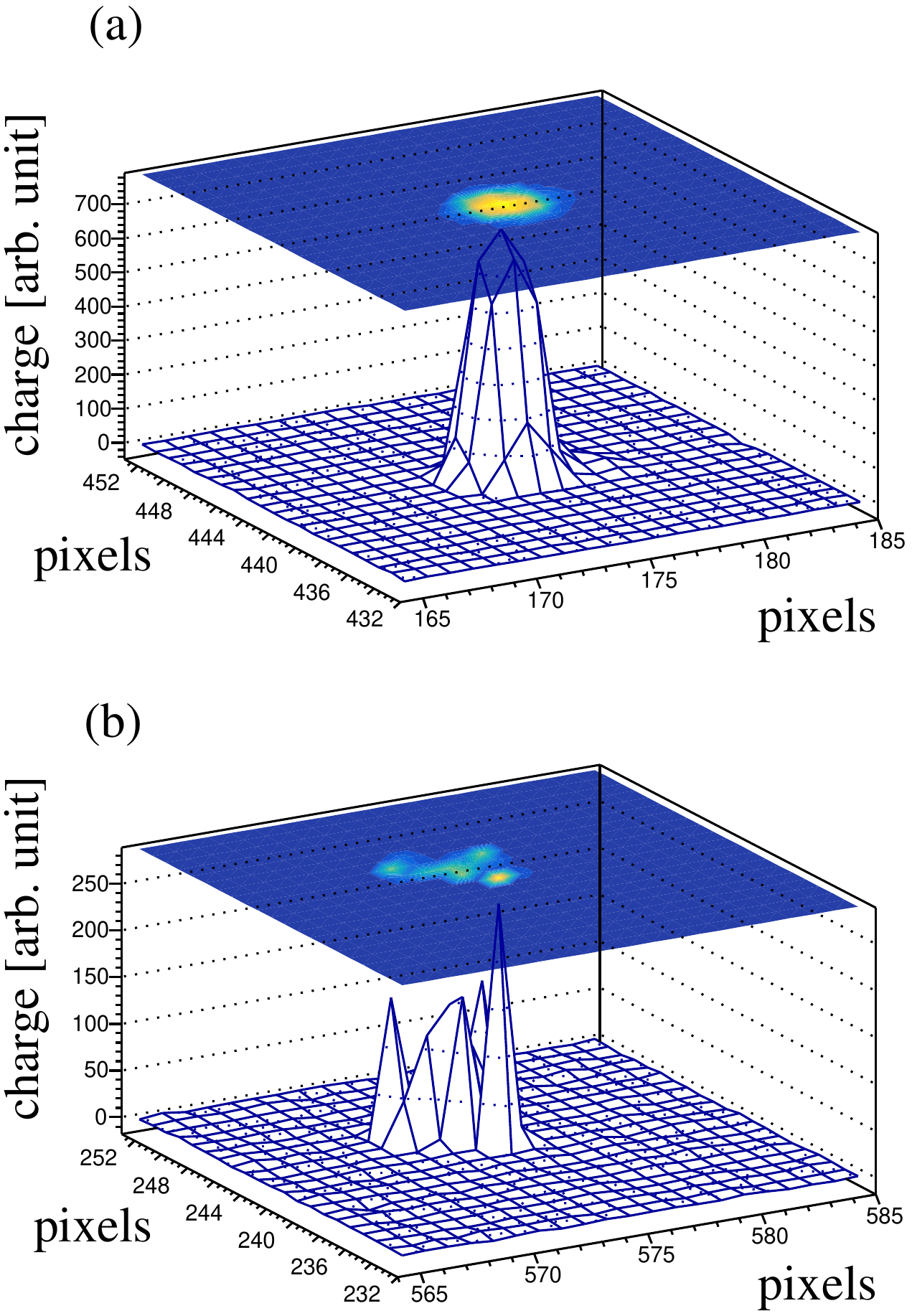}
\caption{
(a) Typical shape of cluster due to heavy charged particle ($\alpha$ or $^6$Li). Symmetric cluster is observed.
(b) Typical image of traveling electron in the active volume. 
These events are expected to be made by the Compton electrons 
from the $\gamma$-ray background.
These clusters show relatively larger $\sigma_y^2$ values.
}
\label{scatt2}
\end{center}
\end{figure}

Fig. \ref{scatt1}(a) shows a scatter plot for $\sigma_y^2$ vs $Q$
(Other identifier $\sigma_x^2$ shows similar distribution to the $\sigma_y^2$).
The profile of these chained mountains in $Q > 0.25$ MeV range was evaluated by fitting with a linear function
$ \sigma_y^2 = aQ + b$ (call it as line L, here after).
The best estimated values were $ \hat{a} = 5.0 \times 10^{-7} $ (pixel)${}^2/$eV and $\hat{b} =  0.39$ (pixel)${}^2$
Projecting data (again in $Q > 0.25$ MeV range) paralelly along the line L to the vertical axis, 
you obtain a $\sigma_y^2$ distribution around the line L, as shown in the figure \ref{scatt1}(b).
Here, we determined the events inside a region $|\sigma_y^2 - \hat{a}Q - \hat{b}| < 0.16$
as neutron-origin events, in which clusters show well symmetric distributions.
The criterion parameter $0.16$ is $3 \sigma$ of a fitted Gausian in the figure \ref{scatt1}(b).
The typical cluster shape for neutron candidate event (around the region A) is shown in the figure \ref{scatt2}(a).
Fig. \ref{scatt2}(b) shows the event around the region B, 
which is expected to represent a traveling electron via Compton scattering with background $\gamma$-rays.

Fig. \ref{scatt1}(c) shows a cluster charge distribution inside the candidate region.
Two large peaks show the $\alpha$ and $^7$Li events from the main branch.
Right side edges of the peaks are shifted 
from where they are supposed to be due solely to the released energies during the nuclear reactions, 
because the secondary particle travels through an inactive volume with a finite length.
The tails in the left sides represent differences in the passing lengths in the inactive volume, 
in which the measured energy varies with the emission angle with respect to the normal incidence.
We fit it to templates of the energy distribution, which made by the Geant4 simulation framework
for the geometry shown in the figure \ref{sensor}, 
with an additional inactive layer on the backside of the sensor.
The thickness of the inactive layer was estimated to be $220 \pm 20$ nm
when one assume that the silicon is a dominant material of it.
The best fit template is shown in the figure \ref{scatt1}(c) as a dashed histogram.
Total number of events measured in this criteria is 4038 for the net measurement time of 7.5 hours,
which corresponds to a detection efficiency of 1.5 \% for the thermal-range neutrons.

\section{Spatial Resolution}

Spatial resolution of the $^{10}$B-INTPIX4 sensor was measured
by evaluating a neutron's shadow, which 
is a projection of the neutron mask made by well-collimated neutron beams, supplied at the BL10 beam line 
of the Materials and Life Science Experimental Facility (MLF), J-PARC center.

\begin{figure}[htb]
\begin{center}
\includegraphics[width=0.85\linewidth]{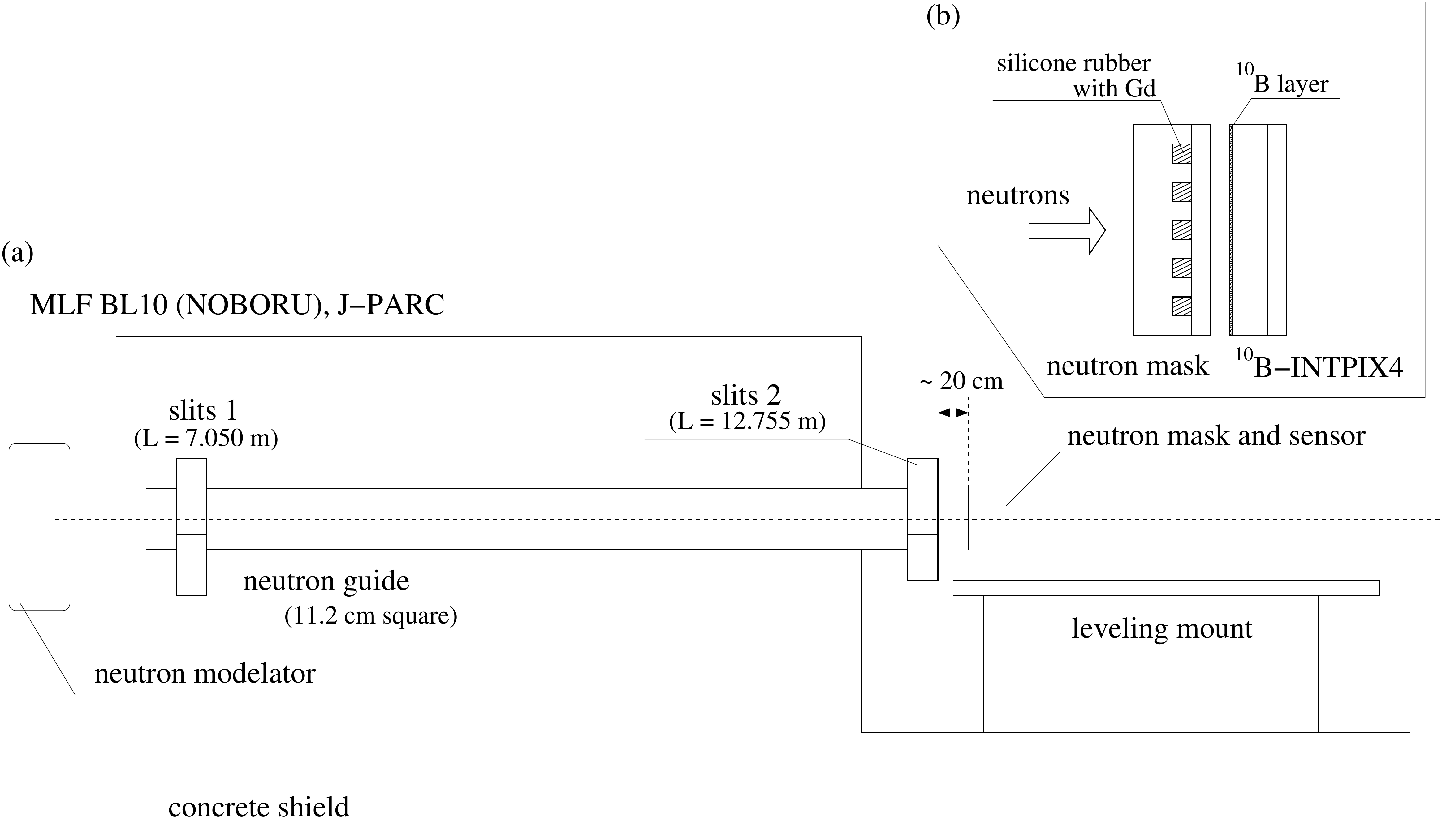}
\caption{(a) Experimental setup for the measurements of spatial resolutions at the BL10 beam line. 
Two B$_4$C collimators (thickness is 5 mm) were set to be 10 mm apertures in vertical and horizontal. 
(b) Closed view of the neutron mask and the $^{10}$B-INTPIX4 sensor.
Distance of the silicone rubber and the $^{10}$B layer was set to be $200 \pm 50 ~\mu$m.
}
\label{setup}
\end{center}
\end{figure}

Fig. \ref{setup}(a) shows schematic drawing of the BL10 beam line.
Two B$_4$C slits are located at 7.050 m and 12.755 m distances
from a surface of a moderator of the MLF, respectively.
We set their opening apertures to be 10 mm squares and selected a small divergence beam of less than 0.1 mrad.
Pb neutron filter with 75 mm thickness was inserted in the beam line to reduce fast neutron components.
Measured transmission of the filter is found in elsewhere \cite{BL10}.

The sensor and the neutron mask is located around 20 cm downstream from the second slits.
Fig. \ref{setup}(b) is a close view of the sensor position.
The neutron mask has a fine structure of the line/space arrangement with $100 ~\mu$m pitch
which is made by chemical etching processes.
Depth of the trenches were $100 ~\mu$m and silicone rubber containing Gd was poured into them, 
in which the neutron beams are partially stopped and makes the shadow on the detector.
Distance between surfaces of the neutron mask and the detector was set to be $200 \pm 50 ~\mu$m.
\begin{figure}[htb]
\begin{center}
\includegraphics[width=0.85\linewidth]{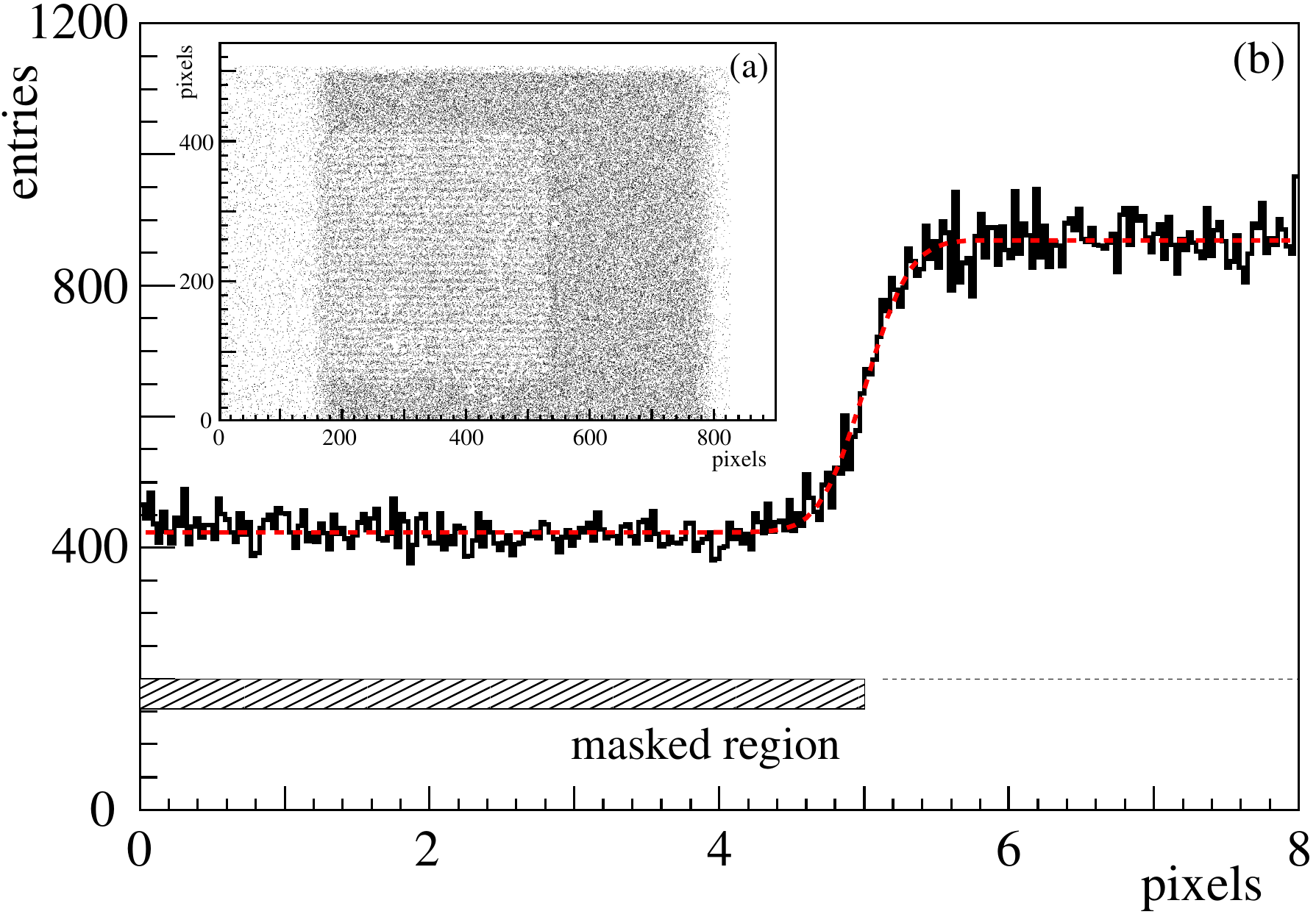}
\caption{
(a) Image of the all area of the sensor (density of dots is reduced for visualization).
Edges of the B$_4$C slits and shadow of the fine neutron mask are observed.
(b) Profile of an accumulated edge shadow.
It is made by superimposing the all edge shadows of the neutron mask.
Spatial resolution is evaluated to be $4.1 \pm 0.2 ~\mu$m,
by fitting with an error function.
}
\label{reso}
\end{center}
\end{figure}

Figure \ref{reso}(a) shows a measured image (density of dots are reduced).
The vertical lines shown at $x=160$ and $x=790$ pixels are due to the beam ends shaped by the slits.
The shadow of the fine neutron mask is imaged around the center.
We chose a region of $180 < x < 520$ and $80 < y < 400$ for the following analysis.
The images in this area of interest is projected to a vertical axis and 
all edge images are superimposed according to the repeated pattern.
Then the histogram of accumulated edges is fitted by an error function
\begin{equation}
A\mathrm{erf}(t) + B \mathrm{, \hspace{2mm} where} ~t = \frac{y-y_0}{\sqrt{2}\sigma_{LSF}}
\end{equation}
with fitting parameters, $A, B, y_0$, and $\sigma_{LSF}$ (the spatial resolution of interest).
The line/space structure is not placed perfectly parallel to the projection direction (horizontal direction in this analysis),
therefore, we rotate the measured image in the x-y plane before the projection, 
and find the minimum $\sigma_{LSF}$ as a function of the rotation angle.
Figure \ref{reso}(b) shows the accumulated shadow edge, 
with the best rotation angle.
We estimate that the upper bound of spatial resolution is to be $4.1 \pm 0.2 ~\mu$m, 
here the uncertainty indicates the fitting error.

\section{Summary}

We have succeeded to measure neutron events by the $^{10}$B-INTPIX4 new neutron imaging sensor.
$\gamma$-ray background in low energy can be rejected by the cluster charge distribution.
Electron events from Compton scattering processes of the higher energy $\gamma$-rays (and cosmic muon events)
are able to be rejected by the additional identifier, the 2nd order center moment effectively.
Detection efficiency with the event criteria is evaluated to be 1.5\% for thermal neutrons.
Upper bound of the spatial resolution is estimated to be $4.1\pm 0.2 ~\mu$m 
as a standard deviation of the line spread function.

\section*{Acknowledgements}
We wish to thank Prof. Takeshi Go Tsuru of Kyoto University 
and all of the members of the SOI group for many supports in the sensor development.
Especially, we are indebted to Dr. Ryutaro Nishimura of KEK for comprehensive developments of 
several tools for readout systems for INTPIX4.
We are deeply grateful to the members of the Takeda Sentanchi super clean room of the Nanotechnology Platform Program
for maintaining the laboratory.

This work was partially supported by JSPS KAKENHI Grant Number 17H05397, 18H04343, and 18H01226.
A part of this work was conducted at Takeda Sentanchi Supercleanroom,
The University of Tokyo, supported by ``Nanotechnology Platform Program" of
the Ministry of Education, Culture, Sports, Science and Technology
(MEXT), Japan, Grant Number JPMXP09F19UT0065.




\bibliography{Hiroshima_Kamiya}

\end{document}